# Curve fitting to resolve overlapping voltammetric peaks: model and examples

W. Huang [a], T.L.E. Henderson [b], A.M. Bond [c,*], K.B. Oldham [d]

[a] *Department of Analytical Chemistry, University of New South Wales, Kensington, NSW 2033, Australia*
[b] *School of Biological and Chemical Sciences, Deakin University, Geelong, Victoria 3217, Australia*
[c] *School of Chemistry, La Trobe University, Bundoora, Victoria 3083, Australia*
[d] *Department of Chemistry, Trent University, Peterborough, Ontario K9J 7B8, Canada*

## Abstract

A model is presented that is applicable to a wide range of peak-shaped voltammetric signals. It may be used, via curve-fitting, to resolve severely overlapped peaks, irrespective of the degree(s) of reversibility of the electrode processes. The resolution procedure has been thoroughly tested for several voltammetric and polarographic techniques (differential pulse, square wave and pseudo-derivative normal pulse), using reversible, quasireversible and irreversible electrochemical systems.

## 1. Introduction

Because the width of a voltammetric peak (typically 100 mV at half height) is an appreciable fraction of the accessible potential range (typically 1500 mV), overlapped peaks occur more commonly in polarography and voltammetry than they do in chromatography or most spectral methods. Accordingly there is a voluminous literature (see Refs. [1–15], and references cited therein, for example) in electroanalytical chemistry on methodologies for resolving conjoined peaks. One successful approach is to use a computer-implemented interactive sequence to fit, to the experimental voltammogram, a curve derived from a theoretical model of the individual electrode processes [8–13]. To date, however, models that are applicable to a wide range of electrode processes, techniques and conditions have not been available. In this paper we shall propose a model that is applicable to a wide variety of chemical systems and electroanalytical techniques; then we shall demonstrate that the curve-fitting method can competently resolve even the severely overlapped peaks that are common in experimental practice. Importantly, no initial guesses of any parameters are required, as is the case with previous popular methods.

## 2. The method

The curve-fitting method requires that a constructed current–voltage relationship, $I_{theor}(E)$, be closely matched to the experimental voltammogram $I_{\text{expt}}(E)$ which arises from the reduction of two analytes that have closely spaced peak potentials (Fig. 1). Of course, the procedure applies equally to oxidations and some of our experiments were oxidations. We regard cathodic current as positive. In principle, extension to three or more overlapping analytes is possible, but we have not pursued that possibility.

---

* Corresponding author.

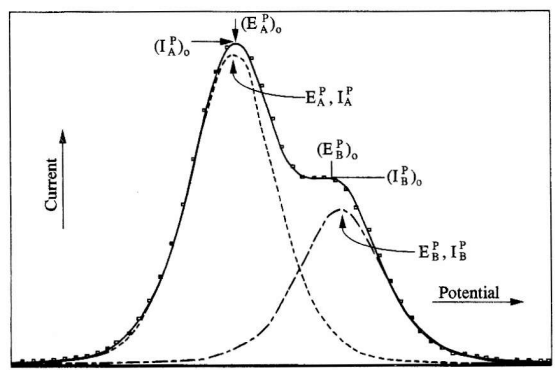

Fig. 1. An example of two overlapping processes and their resolution by curve fitting. The solid line is the simulated curve, the squared line is the fitted curve, and broken lines are the individual peaks.

The fundamental assumption underlying the resolution of overlapped peaks by curve fitting is that the two reductions do not interact in any way, so that

$$I_{\text{expt}}(E) = I_{\text{A alone}}(E) + I_{\text{B alone}}(E) \qquad (1)$$

where $I_{\text{A alone}}(E)$ is the voltammetric curve that would have been produced by analyte A if B had been absent and $I_{\text{B alone}}(E)$ is the voltammogram of analyte B in the absence of A. A more complete description of the voltammetry would include two more terms

$$I_{\text{expt}}(E) = I_{\text{A alone}}(E) + I_{\text{B alone}}(E) + I_{\text{backgd}}(E) + I_{\text{noise}} \qquad (2)$$

where $I_{\text{backgd}}(E)$ is the "background" current–voltage relationship in the absence of both A and B, $I_{\text{noise}}$ being a random signal of indeterminate magnitude and origin. We have assumed, and experiments confirm, that the standard procedure of "blank subtraction" eliminates the effects of $I_{\text{backgd}}(E)$ adequately, although as will be shown later a linear baseline can be tolerated without any problems and this feature can readily be included in the model. Noise has not been explicitly incorporated into our model but simulations (reported briefly later) show that our curve-fitting method can accommodate quite a large admixture of random noise.

In curve fitting, one creates a synthetic current–voltage curve

$$I_{\text{theor}}(E) = I_A(E) + I_B(E) \qquad (3)$$

and progressively improves $I_A(E)$ and $I_B(E)$ in an attempt to minimize (in the least-squares sense) the discrepancy between $I_{\text{theor}}(E)$ and $I_{\text{expt}}(E) - I_{\text{blank}}(E)$. Each of $I_A(E)$ and $I_B(E)$ depends on a number (2, 3 or 4) of parameters via the models which are discussed in the next section. The values of these parameters are varied iteratively until the fit is no longer improved significantly.

Because its basis is well known, the details of our curve-fitting procedure will not be reported here. Suffice it to state that an enhanced version [16] of the "improved [17,18] Marquardt" [19] method was adopted. Starting with initial "guesses" of the (4, 5, 6, 7 or 8) parameters, matrix manipulations within a computer progressively improve those parameter values by minimizing the quantity

$$\sum_{j=1}^{J} \left[ I_{\text{expt}}(E_j) - I_{\text{blank}}(E_j) - I_A(E_j) - I_B(E_j) \right]^2 \qquad (4)$$

where $E_1, E_2, \ldots, E_j, \ldots, E_J$ are the evenly spaced potentials at which the digitized voltammogram (and its corresponding blank) were recorded. After a sufficient number of iterations (typically less than 10), no significant further improvement occurs and the latest batch of parameters is regarded as "correct".

Of course, the assumption is made that the "correct" parameters for peak A are those that make the constructed $I_A(E)$ curve closely match the curve $I_{\text{A alone}}(E) + I_{\text{blank}}(E)$ that would have been obtained experimentally had B been absent. Likewise $I_B(E)$ should match the A-free curve $I_{\text{B alone}}(E) + I_{\text{blank}}(E)$. A post facto confirmation of the efficacy of the resolution is provided by the reconstruction of the entire synthetic voltammogram $I_A(E) + I_B(E) + I_{\text{blank}}(E)$ and its comparison with the experimental $I_{\text{expt}}(E)$ curve.

An early step in the building of a model for a voltammetric peak is to select a set of parameters. We have eschewed a choice based on chemically significant parameters (concentrations, standard potentials, rate constants) in favour of more voltammetrically relevant quantities: the peak coordinates $E^p$ and $I^p$, the reversibility index $\Lambda$, and the symmetry factor $\alpha$. The reversibility index is the dimensionless quantity defined by

$$\Lambda = k^o \sqrt{\pi \Delta t / D} \tag{5}$$

where $k^o$ is the standard heterogeneous rate constant, $D$ is a mean diffusion coefficient, and $\Delta t$ is the characteristic time interval of the voltammetric method, such as a pulse time, drop time or reciprocal frequency. Notice that we are treating the electron number $n$ and the temperature $T$ as known constants. The diffusion coefficient $D$ is not necessarily known; it does not appear explicitly as a parameter, however, being "buried" in the quantities $I^p$ and $\Lambda$.

One advantage of this choice of parameters is that it provides easy and appropriate "guessed" values of $E^p$ and $I^p$ to initialize the Marquardt optimization procedure. Fig. 1 illustrates this. It shows a voltammogram in which the peaks arising from species A and B are severely overlapped. The values marked as $(E_A^p)_o$ and $(I_A^p)_o$ can be used as initial guesses of $E_A^p$ and $I_A^p$, with $(E_B^p)_o$ and $(I_B^p)_o$ being employed similarly. When the voltammograms are so badly merged that only a single peak appears, its location may be used to initialize both $E_A^p$ and $E_B^p$, while its height can be the "guessed" value of both $I_A^p$ and $I_B^p$. The ability to make sensible guesses in this way is very valuable when automatic instrumentation is being used to implement resolution by curve fitting. We used the values 1.000 and 0.500, respectively, as initial guesses of $\Lambda$ and $\alpha$.

## 3. The model

In its most general formulation, the model that we used for a voltammetric peak is

$$I(E) = I^p \frac{f\{\epsilon \sigma\} - f\{\epsilon / \sigma\}}{f\{\sigma\} - f\{1/\sigma\}} \tag{6}$$

The peak current $I^p$ appears as a multiplier, while the peak potential $E^p$ is present within the $\epsilon$ parameter

$$\epsilon = \exp\{(nF/RT)(E^p - E)\} \tag{7}$$

The $\sigma$ term is a known constant

$$\sigma = \exp\{(nF/RT)(\Delta E/2)\} \tag{8}$$

and incorporates the dependence of the voltammetric signal on $\Delta E$, which equals the pulse height in pulse voltammetries, or has an equivalent significance for other techniques. The two kinetic parameters, $\alpha$ and $\Lambda$, are incorporated into the definition, namely

$$f\{x\} = x^\alpha \exp\left\{\frac{\Lambda^2}{\pi}\left[\frac{L+x}{L^\alpha x^{1-\alpha}}\right]^2\right\} \mathrm{erfc}\left\{\frac{\Lambda}{\sqrt{\pi}}\frac{L+x}{L^\alpha x^{1-\alpha}}\right\} \tag{9}$$

of the $f\{x\}$ function. $L$ is used as an abbreviation for

$$L = \left(\frac{\Lambda^3 + \Lambda^2 + \Lambda}{\Lambda^3 + \Lambda^2 + \Lambda + 1}\right)^{1/\alpha} \tag{10}$$

which may be written equivalently as $[(\Lambda^4 - \Lambda)/(\Lambda^4 - 1)]^{1/\alpha}$ but we avoid this succinct representation because it becomes indeterminate when $\Lambda = 1$.

This complicated model is employed in its entirety for quasireversible voltammetric peaks. It may also be employed for voltammograms from species whose degree of reversibility is unknown.

Irreversible voltammograms are characterized by a small value of $\Lambda$. Consequently $L$ then becomes equal to $\Lambda^{1/\alpha}$ and is greatly exceeded by $x$. In these circumstances, the definition of the $f\{x\}$ function simplifies to

$$f\{x\} = x^\alpha \exp\{x^{2\alpha}/\pi\} \mathrm{erfc}\{x^\alpha/\sqrt{\pi}\} \text{ (irreversible)} \tag{11}$$

so that the dependence on the $\Lambda$ parameter disappears.

Conversely, reversibility corresponds to a large value of $\Lambda$. In that limit, $L$ equals unity and $f\{x\}$ therefore acquires the form

$$x^\alpha \exp\{\Lambda^2(1+x)^2/(\pi x^{2-2\alpha})\}$$
$$\times \mathrm{erfc}\{\Lambda(1+x)/(\sqrt{\pi} x^{1-\alpha})\}$$

However, when, as here, its argument is large, the $\exp\{z^2\}\mathrm{erfc}\{z\}$ product becomes equal to $1/\sqrt{\pi}z$, so that

$$f\{x\} = \frac{x}{\Lambda(1+x)} \quad \text{(reversible)} \tag{12}$$

Moreover, Eq. 6 then simplifies to

$$I(E) = I^{\mathrm{p}} \frac{\epsilon(1+\sigma)^2}{(\epsilon+\sigma)(\epsilon\sigma+1)} \quad \text{(reversible)} \tag{13}$$

Thus the model for a reversible peak has only two unknown parameters: the peak current $I^{\mathrm{p}}$ and (via $\epsilon$) the peak potential $E^{\mathrm{p}}$.

In constructing our model, we leaned heavily on the theory of pseudo-derivative normal pulse polarography [20-23]. For that technique, Eq. 13 is known to be exact under reversible conditions. Similarly the combination of Eqs. 6 and 11, though incorporating small errors otherwise, is obeyed by irreversible pseudo-derivative normal pulse polarography when $\Delta E$ is sufficiently small. Eq. 9 itself is empirical; it was selected because it reduces appropriately to Eq. 11 or 12 for extreme values of the reversibility index $\Lambda$, is consistent with theory for the small amplitude case available with some techniques [24] and because it provides an excellent numerical match to simulated quasireversible pseudo-derivative normal pulse polarograms.

Despite its origins in the pseudo-derivative technique, there are good reasons for the belief that the model should also fit many other electroanalytical methods that generate peak-shaped current-voltage curves [25]. In the subsequent sections of this paper, we shall describe successful applications to voltammograms (at glassy carbon and stationary mercury electrodes) and polarograms (at dropping mercury electrodes) generated by three electroanalytical techniques: differential pulse, pseudo-derivative normal pulse, and square wave. The approach should also be applicable to semidifferentiated linear-potential-sweep voltammetry, a.c. polarography and other techniques.

## 3.1. Background current

If a baseline current $I_{\mathrm{b}}$ is a linear function of potential $E$, which is a good approximation over a narrow potential range, then

$$I_{\mathrm{b}} = a + bE \tag{14}$$

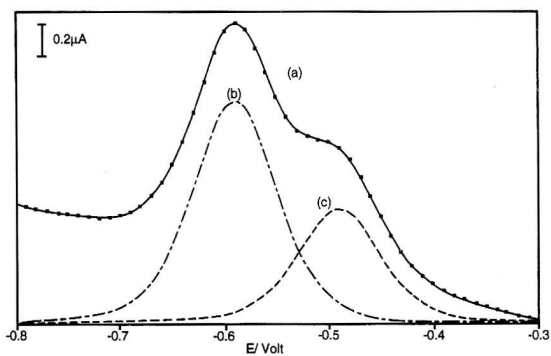

Fig. 2. Resolution of overlapping processes with a linear baseline. The solid line (curve) is the simulated curve with data represented by squares being obtained by curve fitting and broken lines (curves b, c) being the resolved individual process. Parameters used to simulate curve (a) are $n_1 = n_2 = 1$, $c_1 = 1 \times 10^{-4}$ M, $c_2 = 2 \times 10^{-4}$ M, $\Delta E_{1/2} = -100$ mV, $T = 298.2$ K, $A = 0.01$ cm$^2$, $D_1 = D_2 = 4 \times 10^{-6}$ cm$^2$ s$^{-1}$, $t_{\mathrm{p}} = 0.045$, $E_{\mathrm{step}} = 1$ mV, $E_{1/2,1} = -500$ mV.

where $a$ is an intercept and $b$ is a slope. Unfortunately, it is difficult to generate the initial guesses of $a$ and $b$ via direct use of this equation. However, Eq. 15 may be used for automatic initialization of these parameters

$$I_{\mathrm{b}} = I_{\mathrm{i}} + \frac{(E - E_{\mathrm{i}})(I_{\mathrm{f}} - I_{\mathrm{i}})}{(E_{\mathrm{f}} - E_{\mathrm{i}})} \tag{15}$$

where $E_{\mathrm{i}}$ is an initial potential, $E_{\mathrm{f}}$ is a final potential, $I_{\mathrm{i}}$ is the current $I(E_{\mathrm{i}})$ at the first point and $I_{\mathrm{f}}$ is the current $I(E_{\mathrm{f}})$ at final point. $I_{\mathrm{i}}$ and $I_{\mathrm{f}}$ are now taken as the unknown model parameters to be optimized.

This background current equation can be combined with the model for fitting the polarograms with an assumed linear baseline current. Computer simulations (Fig. 2) illustrate that a linear background current has no effect on the deconvolution of overlapping reversible processes when Eq. 15 is incorporated into the reversible model. Further details on this process are available in Ref. [16].

## 3.2. Random noise

Practical polarograms contain some noise which is often, but not always, random in character. The background current considered above, would be described as systematic or non-random noise.

To evaluate the influence of random noise in applying curve fitting methods, the noise may be added to the Faradaic current response point by point so as to mimic the practical situation

$$I_t = I_{\text{Far},j} + I_{\text{noise},j} \qquad (16)$$

Random noise is, of course, not correlated with the values of the Faradaic current nor the values of the corresponding abscissae (i.e., the potential) at each point. A simulated noisy polarogram (Fig. 3) is therefore produced by adding Gaussian noise to a noise-free polarogram [16,27]. In Fig. 3, the value of the signal-to-noise ratio is 10 and the Faradaic response clearly is visible. However, if the amplitude of the Faradaic signal is less than or equal to the noise, i.e., $S/N \leq 1$, the Faradaic signal is masked by the noise. Data in Table 1 demonstrate that random noise has no marked effect on resolving the signals of individual components from overlapping processes with $S/N = 20$ and 10 by the curve fitting procedure, although a $S/N$ ratio of 5 has a noticeable effect.

### 3.3. Chemical systems used

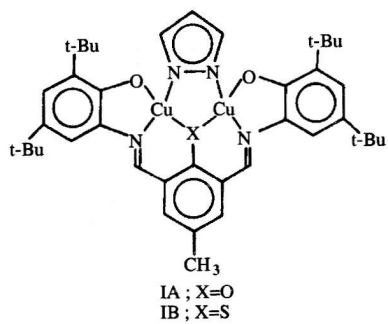

IA ; X=O
IB ; X=S

Our first examples of overlapping processes derive from binuclear copper complexes having structure **I** where X = S or O. The two binuclear complexes differ only in the nature of the endogenous bridging atom, and may be represented as L(S)Cu$_2$(pz) and L(O)Cu$_2$(pz), where pz symbolises the pyrazolate bridge and L represents the remaining part of the ligand system. A sequence of one-electron charge-transfer processes have been reported for each compound [26]. However, a major problem in studying these binuclear complexes was that strongly overlapping responses made it difficult (or impossible) to determine the required electrochemical parameters accurately. This paper shows how the curve-fitting method may be successfully applied to solve this problem.

The second class of overlapping processes studied here derives from mixtures of In(III) + Cd(II) and Tl(I) + Pb(II). Both In(III) and Cd(II) are reversibly reduced in aqueous hydrochloric acid at mercury electrodes according to the processes

$$\text{In(III)} + 3e^- \rightarrow \text{In(Hg)} \qquad (16)$$

$$\text{Cd(II)} + 2e^- \rightarrow \text{Cd(Hg)} \qquad (17)$$

The In(III), Cd(II) system is regarded as a classical overlapping response problem in polarography [2,3,11,12] with the reversible half-wave potentials separated by about 40 mV in 1.0 M HCl. The reductions of Pb(II) and Tl(I) at mercury electrodes to their amalgams also are reversible and occur as in Eqs. 18 and 19, respectively

$$\text{Pb(II)} + 2e^- \rightarrow \text{Pb(Hg)} \qquad (18)$$

$$\text{Tl(I)} + e^- \rightarrow \text{Tl(Hg)} \qquad (19)$$

This problem also represents a well-known example of two species giving rise to severely overlapping polarographic signals [3,27–29]. For example, their peak potentials in differential pulse polarography are separated by about 60 mV in 1.0 M ZnSO$_4$ acidified to pH 2 with H$_2$SO$_4$ [27].

The example represented by Eqs. 16 and 17 corresponds to a case in which a three-electron reversible process overlaps with a reversible two-electron charge-transfer process. In contrast, the lead(II) and thallium(I) system involves a reversible two-electron process overlapping with a reversible one-electron charge-transfer process and the binuclear copper complexes provide reversible one-electron overlapping responses. Reversible one-electron processes are considerably broader than two or three-electron charge-transfer steps so the above examples represent a severe test of overlapping responses constructed from different combinations of shapes.

The third and final category of reactions considered concerns the irreversible reduction of chromium and the quasireversible reduction of zinc ions at mercury electrodes. These two species exhibit the

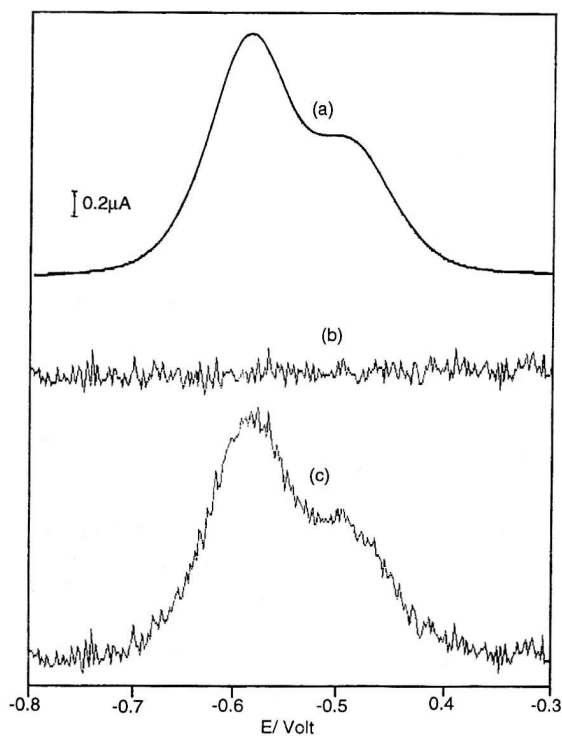

Fig. 3. An example of overlapping processes in the presence of noise where curve c represents the observed response calculated as the sum of the noise free response (curve a) and the random noise (curve b). $S/N = 10$, other parameters are as for Fig. 2.

following overall reactions at a mercury electrode [30]:

$$Cr(III) + e^- \rightarrow Cr(II) \quad (20)$$

$$Zn(II) + 2e^- \rightarrow Zn(Hg) \quad (21)$$

Therefore, the Cr(III)/Cr(II) reaction is one-electron charge transfer process with the product of reaction, Cr(II), being soluble in aqueous solution. In contrast, the Zn(II)/Zn(Hg) reduction involves a two-electron charge transfer and the product, elemental zinc, forms an amalgam with the mercury electrode. A mixture of Cr(III) and Zn(II) therefore provides a good example of overlapping irreversible, solute-forming, and quasireversible, amalgam-forming, processes. The individual processes for reduction of chromium and zinc ions have been the subjects of extensive studies [24,30–32]. However, the overlapping responses of these two species under conditions of differential pulse polarography and pseudo-derivative normal pulse polarography have not been reported, although data on overlapping DC polarographic waves for reduction of chromium and zinc ions are available [12].

The Cr(III) and Zn(II) system, as well as providing an opportunity to assess the general model for resolution of overlapping kinetically controlled processes, also provides a thorough test of our model.

## 4. Experimental

*Chemicals*

Unless otherwise specified, all chemicals used were of analytical grade purity. The binuclear complexes $L(S)Cu_2(pz)$ and $L(O)Cu_2(pz)$ were prepared as described in Ref. [26] and dried under vacuum for 8 h.

Table 1
Effect of random noise on the resolution of overlapping reversible processes [a]

| $S/N$ | Peak | Expected | | Calculated | | Error [b] | |
|---|---|---|---|---|---|---|---|
| | | $I^p$ ($\mu A$) | $E^p$ (mV) | $I^p$ ($\mu A$) | $E^p$ (mV) | $RE(I^p)$ (%) | $AE(E^p)$ (mV) |
| 20 | 1 | 1.046 | −490 | 1.091 | −488.7 | 4.3 | 1.3 |
| | 2 | 2.092 | −590 | 2.155 | −589.8 | 3.0 | 0.2 |
| 10 | 1 | 1.046 | −490 | 1.143 | −487.4 | 9.3 | 2.6 |
| | 2 | 2.092 | −590 | 2.227 | −589.6 | 6.5 | 0.4 |
| 5 | 1 | 1.046 | −490 | 1.261 | −484.5 | 20.1 | 5.5 |
| | 2 | 2.092 | −590 | 2.376 | −589.7 | 13.6 | 0.3 |

[a] $n_1 = n_2 = 1$, $c_1 = 1 \times 10^{-4}$ M, $c_2 = 2 \times 10^{-4}$ M, $E^r_{1/2,1} = 500$ mV, $\Delta E^r_{1/2} = 100$ mV, $T = 298.2$ K, $A = 0.01$ cm$^2$, $D_1 = D_2 = 4 \times 10^{-6}$ cm$^2$ s$^{-1}$, $t_p = 0.04$ s, $E_{step} = 1$ mV.
[b] $RE(I^p_j) = I^p_{cal,j}/I^p_{exp,j} - 1$ and $AE(E^p_j) = E^p_{cal,j} - E^p_{exp,j}$.

To prepare the required metal ion solutions in aqueous media, appropriate volumes of standard solutions of $Cd(NO_3)_2$, $In(NO_3)_3$, $Pb(NO_3)_2$, $TlNO_3$, $Cr(NO_3)_3$ and $Zn(NO_3)_2$ supplied by BDH, were added to the electrolyte.

For studies with the binuclear complexes, HPLC-grade dichloromethane and electrochemical grade tetrabutylammonium tetrafluoroborate electrolyte purchased from Southwestern Analytical Chemicals (Austin, TX), were used, after drying under vacuum for at least 8 h.

*Instrumentation and procedures*

All solutions for voltammetric studies were deoxygenated for 5 min with high purity nitrogen prior to experiments. Data are reported for a temperature of $25 \pm 1°$C.

Polarographic data for the study of the overlapping indium and cadmium processes were obtained with a Metrohm Model 646 Voltammetric Analysis Processor and a Metrohm Model 647 electrode assembly. The Metrohm Model 647 contains a multi-mode mercury working electrode [which may be used as a dropping mercury electrode (DME), a static mercury drop electrode (SMDE) or a hanging mercury drop electrode (HMDE)], an aqueous Ag/AgCl (saturated KCl) reference electrode and a glassy carbon auxiliary electrode.

Polarograms for the overlapping lead and thallium processes were obtained with a BAS100 electrochemical analyzer used in conjunction with a PAR Model 303 static mercury drop electrode system (SMDE or HMDE modes), a Ag/AgCl (saturated KCl) reference electrode and a platinum auxiliary electrode.

Voltammetric measurements on the binuclear copper complexes were performed with the BAS-100 electrochemical analyzer at platinum disk (Pt), gold disk (Au) and glassy carbon disk (GC) electrodes. The solid working electrodes were polished frequently with an aqueous alumina slurry, followed by rinsing with water and then with solvent ($CH_2Cl_2$). This cleaning treatment was required to obtain reproducible results. The auxiliary electrode was a platinum wire and the reference electrode was Ag/AgCl ($CH_2Cl_2$; saturated LiCl). Potentials in dichloromethane are referenced to the potential of the Fc/$Fc^+$ process [Fc = ferrocene, $(\eta^5\text{-}C_5H_5)_2Fe$] via measurement of the reversible potential versus Ag/AgCl for oxidation of a $5 \times 10^{-4}$ M solution of ferrocene in dichloromethane.

The experimental data were transferred from the Metrohm 646 VA Processor or the BAS-100 Analyzer to a SPHERE microcomputer and then either to a mainframe DEC-20 computer or to an IBM personal computer for processing.

Unless otherwise stated, the experimental conditions used for voltammetric measurements with the BAS-100 equipment were:
(i) Normal and differential pulse voltammetry (polarography): drop time or duration between pulses 1 s; pulse time 0.06 s, current sampling time 0.02 s; pulse amplitude $-50$ mV; scan rate $-2$ mV s$^{-1}$; potential step $-2$ mV.
(ii) Square-wave voltammetry (polarography): drop time 1 s; square-wave amplitude 25 mV; square-wave frequency 12 Hz; potential step $-2$ mV.

The measurement parameters used with the Metrohm equipment, except where stated, were as follows:
(i) Normal and differential pulse voltammetry (polarography): drop time or duration between pulses 0.6 s; pulse amplitude $-50$ mV; pulse time 0.04 s, current sampling time 0.02 s; scan rate $-2$ mV s$^{-1}$.
(i) Square-wave voltammetry (polarography): drop time 0.6 s; square-wave amplitude 25 mV; scan rate $-2.5$ mV s$^{-1}$.

For the studies on the binuclear copper complexes in dichloromethane, the concentrations of L(S)Cu$_2$(pz), L(O)Cu$_2$(pz) and ferrocene were all $5 \times 10^{-4}$ M, the potential step was $+4$ mV, the duration between pulses was 1 s, the potential range examined was from 200 to 1200 mV vs. Ag/AgCl, the scan rate was 4 mV s$^{-1}$, and other parameters are as for the above mentioned experiments with the BAS instrument.

## 5. Results and discussion

*Oxidation of binuclear copper complexes*

For the binuclear, L(X)Cu$_2$(pz), complexes the processes studied [26] were the oxidation steps:

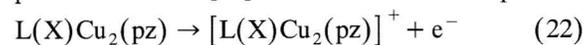

$$L(X)Cu_2(pz) \rightarrow [L(X)Cu_2(pz)]^+ + e^- \qquad (22)$$

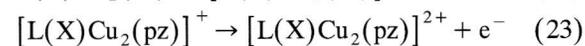

$$[L(X)Cu_2(pz)]^+ \rightarrow [L(X)Cu_2(pz)]^{2+} + e^- \qquad (23)$$

Table 2
Resolution of the two overlapping reversible processes by the curve fitting method applied to pseudo-derivative normal pulse voltammetric data for oxidation of $5 \times 10^{-4}$ M L(O)Cu$_2$(pz) in CH$_2$Cl$_2$ (0.1 M Bu$_4$NBF$_4$) at a glassy carbon electrode as a function of the pulse amplitude [a]

| $\Delta E$ (mV) | $E_{1/2,1}^{rev}$ (mV) [b] | $I_1^p$ ($\mu$A) | $E_{1/2,2}^{rev}$ (mV) [b] | $I_2^p$ ($\mu$A) | $\Delta E_{1/2}^{rev}$ (mV) | $I_2^p/I_1^p$ | R.S.D. [c] (%) |
|---|---|---|---|---|---|---|---|
| 8 | 399 | 7.15 | 443 | 6.90 | 44 | 0.965 | 1.45 |
| 20 | 399 | 17.8 | 444 | 17.2 | 45 | 0.964 | 1.30 |
| 32 | 399 | 27.6 | 443 | 27.5 | 44 | 0.997 | 1.24 |
| 48 | 398 | 38.7 | 443 | 41.1 | 45 | 1.06 | 1.12 |

[a] Potential interval 4 mV; duration between pulses 1 s; scan rate 4 mV s$^{-1}$; pulse width 0.06 s; current sampling width 0.02 s.
[b] mV vs. Fc/Fc$^+$.
[c] Relative standard deviation = $[\Sigma(I_{cal} - I_{exp})^2/N]^{1/2}/I_{max}$. For a perfect fit, R.S.D. would be zero.

where X = O or S. Theoretically, for these systems, equal peak heights should be achieved for all experiments as both processes are derived from oxidation of the same species.

A differential pulse voltammogram for the oxidation of L(S)Cu$_2$(pz) and Fc in CH$_2$Cl$_2$ (0.10 M Bu$_4$NBF$_4$) at a platinum disk electrode shows three peaks, labelled I, II and III in Fig. 4a where peak II strongly overlaps peak III. Peak I is the Fc oxidation peak used as an internal standard, and peaks II and III correspond to processes 22 and 23, respectively.

In contrast, it is not possible to visually distinguish the two oxidation steps for L(O)Cu$_2$(pz) from the pseudo-derivative normal pulse voltammogram of L(O)Cu$_2$(pz) at a glassy carbon electrode (Fig. 4b). The difference between the two half-wave potentials for L(O)Cu$_2$(pz) ($\Delta E_{1/2}^{rev}$) therefore must be smaller than that for L(S)Cu$_2$(pz). However, use of our model for resolving two overlapping reversible processes gives the data presented in Tables 2 and 3 for pseudo-derivative normal pulse voltammetry and for differential pulse voltammetry in Table 4. The $E_{1/2}^{rev}$ values in Tables 2 to 4 are calculated from the relationship $E_{1/2}^{rev} = E_p - \Delta E/2$.

Table 3
Resolution of the two overlapping reversible processes by the curve fitting method applied to pseudo-derivative normal pulse voltammetric data for oxidation of $5 \times 10^{-4}$ M L(S)Cu$_2$(pz) in CH$_2$Cl$_2$ (0.1 M Bu$_4$NBF$_4$) at different electrodes and as a function of pulse amplitude [a]

| Electrode | $\Delta E$ (mV) | $E_{1/2,1}^{rev}$ (mV) [b] | $I_1^p$ ($\mu$A) | $E_{1/2,2}^{rev}$ (mV) [b] | $I_2^p$ ($\mu$A) | $\Delta E_{1/2}^{rev}$ (mV) | $I_2^p/I_1^p$ | R.S.D. (%) |
|---|---|---|---|---|---|---|---|---|
| Au | 8 | 340 | 0.618 | 415 | 0.698 | 75 | 1.13 | 1.23 |
|    | 20 | 340 | 1.53 | 415 | 1.73 | 75 | 1.13 | 0.967 |
|    | 32 | 340 | 2.41 | 415 | 2.73 | 75 | 1.13 | 0.891 |
|    | 48 | 340 | 3.46 | 415 | 3.93 | 75 | 1.14 | 0.778 |
|    | 100 | 342 | 5.58 | 413 | 6.55 | 71 | 1.17 | 1.52 |
| Pt | 8 | 336 | 0.742 | 422 | 0.613 | 86 | 0.827 | 3.02 |
|    | 20 | 336 | 1.84 | 422 | 1.52 | 86 | 0.830 | 2.94 |
|    | 32 | 335 | 2.88 | 422 | 2.40 | 87 | 0.834 | 2.80 |
|    | 48 | 335 | 4.11 | 422 | 3.47 | 87 | 0.843 | 2.41 |
|    | 100 | 336 | 6.76 | 421 | 5.80 | 85 | 0.857 | 0.948 |
| GC | 8 | 350 | 7.26 | 426 | 6.72 | 76 | 0.926 | 3.43 |
|    | 20 | 350 | 18.1 | 426 | 16.6 | 76 | 0.916 | 3.23 |
|    | 32 | 350 | 29.0 | 427 | 26.1 | 77 | 0.901 | 2.83 |
|    | 48 | 351 | 42.9 | 428 | 37.6 | 77 | 0.877 | 2.27 |
|    | 100 | 352 | 70.0 | 426 | 65.8 | 74 | 0.940 | 1.81 |

[a] Parameters not specified are as given in Table 2.
[b] mV vs. Fc/Fc$^+$.

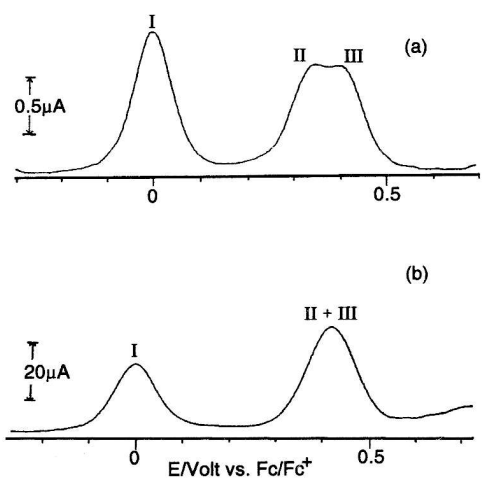

Fig. 4. Pulse voltammograms for oxidation of $5 \times 10^{-4}$ M each of ferrocene (process I) and $L(X)Cu_2(pz)$ (processes II and III) in dichloromethane (0.10 M $Bu_4NBF_4$). (a) Differential pulse voltammograms for oxidation of $L(S)Cu_2(pz)$ at a platinum electrode. $\Delta E = -10$ mV. (b) Pseudo-derivative normal pulse voltammogram for oxidation of $L(O)Cu_2(pz)$ at a glassy carbon electrode. $\Delta E = -20$ mV.

Table 2 shows that, as theoretically expected, the half-wave potentials for the oxidation of $L(O)Cu_2(pz)$ at a glassy carbon electrode are independent of pulse amplitude. The two peak currents, $I_1^p$ and $I_2^p$, increase as the pulse amplitude increase, but their ratio remains close to unity, again as theoretically expected. Furthermore, the relative standard deviation, a criterion of the goodness of fit, is very small. Results close to those theoretically expected also are found for $L(S)Cu_2(pz)$ (Tables 3 and 4) and confirm that the curve-fitting method may be usefully applied for the resolution of overlapping reversible processes, using either pseudo-derivative or differential pulse voltammetry at solid electrodes.

In the case of $L(S)Cu_2(pz)$, a slight dependence of the calculated half-wave potentials on electrode material is evident under conditions of pseudo-derivative normal pulse voltammetry. However, this is less evident with the differential pulse voltammetric method. The variation of calculated $E_{1/2}^{rev}$ with the pseudo-derivative normal pulse voltammetric method is mainly attributable to the presence of a non linear baseline, which is electrode dependent and therefore not accounted for by the model, which accommodates a linear baseline only. In the differential pulse method, a relatively flat baseline is observed and $\Delta E_{1/2}^{rev}$ is calculated to be 75 mV at a gold electrode, 81 mV at a platinum electrode and 79 mV at a glassy carbon electrode. These $\Delta E_{1/2}^{rev}$ values are consistent with the estimated literature value of 80–90 mV [26].

The data for oxidation of $L(X)Cu_2(pz)$ show that curve fitting methods based on deconvolution of reversible overlapping processes are highly reliable by either the pseudo-derivative normal pulse or differential pulse voltammetric techniques provided the baseline is linear, as required by the model. In the case of the oxidation of $L(X)Cu_2(pz)$ the non-linear baseline causes some departure from ideality in the deconvolution polarogram.

*In(III) + Cd(II) system*

Fig. 5 shows that the fitting of an experimental differential pulse polarogram obtained for a mixture

Table 4
Resolution of two overlapping reversible processes by the curve fitting method applied to differential pulse voltammetric data for oxidation of $5 \times 10^{-4}$ M $L(S)Cu_2(pz)$ in $CH_2Cl_2$ (0.1 M $Bu_4NBF_4$) at different electrodes and as a function of pulse amplitude [a]

| Electrode | $\Delta E$ (mV) | $E_{1/2,1}^{rev}$ (mV) [b] | $I_1^p$ ($\mu$A) | $E_{1/2,2}^{rev}$ (mV) [b] | $I_2^p$ ($\mu$A) | $\Delta E_{1/2}^{rev}$ (mV) | $I_2^p/I_1^p$ | R.S.D. (%) |
|---|---|---|---|---|---|---|---|---|
| Au | 10 | 336 | 0.846 | 411 | 0.892 | 75 | 1.05 | 1.99 |
|    | 50 | 336 | 3.92 | 410 | 3.87 | 74 | 0.986 | 0.522 |
| Pt | 10 | 335 | 0.744 | 416 | 0.745 | 81 | 1.00 | 1.46 |
|    | 50 | 338 | 3.65 | 419 | 3.45 | 81 | 0.940 | 0.877 |
| GC | 10 | 331 | 8.77 | 410 | 8.73 | 79 | 0.995 | 1.29 |
|    | 50 | 337 | 38.5 | 415 | 33.9 | 78 | 0.881 | 0.767 |

[a] Parameters not specified are as given in Table 2 or the Experimental section.
[b] mV vs. $Fc/Fc^+$.

10                          W. Huang et al. / Analytica Chimica Acta 304 (1995) 1–15

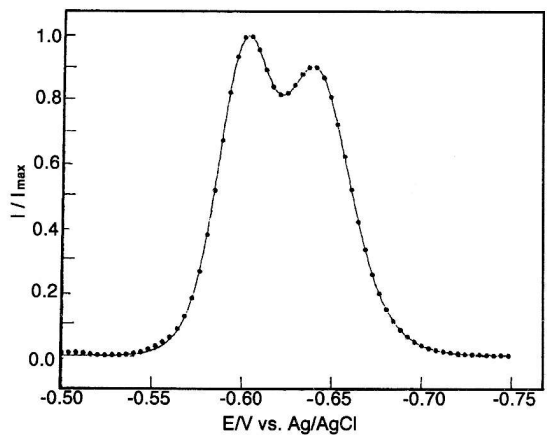

Fig. 5. Comparison of experimental ($\cdots$) and calculated (solid line) differential pulse polarograms with $\Delta E = -10$ mV for reduction of a mixture of $1.74 \times 10^{-5}$ M In(III) and $3.56 \times 10^{-5}$ M Cd(II) in 1.0 M HCl. The current axis has been normalized by division by the maximum current $I_{max}$.

of In(III) and Cd(II) in 1.0 M HCl at a small pulse amplitude ($\Delta E = -10$ mV) is excellent since the residual is very small. However, the fit is worse in differential pulse polarography when a large pulse amplitude is used because a significant d.c. component is introduced into the experiment which is not eliminated in any subtraction process [33]. Thus, Fig. 6a illustrates that fits to the experimental data with a large pulse amplitude ($\Delta E = 100$ mV) are not as good as for the small amplitude curve (Fig. 5). The discrepancies caused by the d.c. term in Fig. 6a is most noticeable around the peak and at the negative potential region of the polarogram. If it is assumed that a linear baseline problem is present then the problem of the d.c. offset can be minimised (Fig. 6b). Table 5 presents data obtained by the differential pulse method at a DME and HMDE and square wave polarography at the DME. Table 6 lists results obtained by the different methods as a function of the concentration ratio of In(III) to Cd(II).

Table 5
Data obtained for resolution of overlapping In(III) and Cd(II) processes in 1 M HCl (concentrations: In(III), $1.74 \times 10^{-5}$ M; Cd(II), $3.56 \times 10^{-5}$ M)

(A) differential pulse polarography at the DME and differential pulse voltammetry at the HMDE [a]

| Electrode | $\Delta E$ (mV) | Species | Expected [b] | | Calculated [c] | |
|---|---|---|---|---|---|---|
| | | | $I^p$ ($\mu$A) | $E^p$ (mV) | $I^p$ ($\mu$A) | $E^p$ (mV) |
| DME | $-10$ | In(III) | 0.195 | $-598$ | 0.1953 | $-598$ |
| | | Cd(II) | 0.185 | $-638$ | 0.1850 | $-640$ |
| | $-100$ | In(III) | 0.704 | $-598$ | 0.7118 | $-599$ |
| | | Cd(II) | 0.976 | $-638$ | 0.9410 | $-638$ |
| HMDE | $-10$ | In(III) | 0.0766 | $-598$ | 0.07401 | $-599$ |
| | | Cd(II) | 0.0872 | $-642$ | 0.09120 | $-643$ |
| | $-100$ | In(III) | 0.260 | $-598$ | 0.2605 | $-599$ |
| | | Cd(II) | 0.398 | $-640$ | 0.3769 | $-642$ |

(B) Square wave polarography at DME [d]

| | $\Delta E$ (mV) | Species | Experimental | | Calculated | |
|---|---|---|---|---|---|---|
| | | | $I^p$ ($\mu$A) | $E^p$ (mV) | $I^p$ ($\mu$A) | $E^p$ (mV) |
| | 5 | In(III) | 0.0768 | $-598$ | 0.07760 | $-597$ |
| | | Cd(II) | 0.0825 | $-638$ | 0.08251 | $-640$ |
| | 25 | In(III) | 0.352 | $-598$ | 0.3515 | $-597$ |
| | | Cd(II) | 0.382 | $-638$ | 0.3871 | $-639$ |

[a] Experimental parameters are: drop time, 1 s; potential step, 2 mV; scan rate, 2 mV s$^{-1}$; pulse width, 0.04 s; current sampling width, 0.02 s.
[b] Values obtained from experimental measurements on single component solutions.
[c] Values obtained on mixtures of the two components and use of the model developed in this paper to resolve overlapping voltammetric processes.
[d] Parameters not stated are defined in the text.

For the overlapping In(III) + Cd(II) system in 1.0 M HCl, two partially overlapped peaks are observed if $\Delta E$ is 10 mV (see Fig. 5), but only one peak appears when $\Delta E$ is 100 mV (Fig. 6). However, as can be seen from Table 5, excellent recovery of the expected values is obtained even when $\Delta E$ is as large as 100 mV. The expected values in Table 5 and elsewhere were determined from data obtained from measurements on single component systems. It should be noted that for differential pulse polarography with the BAS 100 instrument, the potential axis is plotted on the basis of the potential ($E_1$) before the pulse so the peak potential is a function of $\Delta E$. However, with the Metrohm instrument, the peak position of individual electroactive species is independent of the pulse amplitude as $(E_1 + E_2)/2$ rather than $E_1$ is used as the basis of the potential axis, $E_2$ being the potential after the pulse.

The concentration ratio also affects the shape of the overlapping processes. For the In(III) + Cd(II) system in 1.0 M HCl, when the concentration ratio of In(III) to Cd(II) [$c(Cd)/c(In)$] is less than 1, a single peak is observed and the apparent peak potential is very close to the peak potential of In(III). When $c(Cd)/c(In)$ is near 2, two peaks are observed, but when $c(Cd)/c(In) > 3$, a single peak reappears. However, under these conditions the apparent peak potential is close to the peak potential of Cd(II). Despite the peak potential and shape change, data in Table 5 indicate that the curve-fitted results for resolution of overlapping differential pulse and square wave processes at both the DME and HMDE for the concentration ratios of $c(Cd)/c(In)$ from 0.5 to 10 are in excellent agreement with the theoretically expected results. Data in Table 6 also confirm that the curve-fitting method for reversible processes can be applied for different values of $\Delta E$, different concentration ratios, different techniques and different forms of mercury electrode.

Table 6
Effect of concentration ratio on the resolution of overlapping processes for a mixture of In(III) and Cd(II) in 1 M HCl with different voltammetric techniques and pulse amplitudes [a]

| Electrode | $c(Cd)/c(In)$ | Species | Expected | | Calculated | |
|---|---|---|---|---|---|---|
| | | | $I^p$ ($\mu$A) | $E^p$ (mV) | $I^p$ ($\mu$A) | $E^p$ (mV) |
| (A) Differential pulse polarography at the DME and differential pulse voltammetry at the HMDE | | | | | | |
| DME | 0.511 | In(III) | 0.710 | −594 | 0.6991 | −591 |
| | | Cd(II) | 0.191 | −636 | 0.1920 | −632 |
| | 2.04 | In(III) | 0.362 | −594 | 0.3574 | −596 |
| | | Cd(II) | 0.370 | −636 | 0.3559 | −637 |
| | 10.2 | In(III) | 0.362 | −594 | 0.3698 | −595 |
| | | Cd(II) | 1.82 | −636 | 1.815 | −636 |
| HMDE | 0.511 | In(III) | 0.312 | −592 | 0.3237 | −592 |
| | | Cd(II) | 0.100 | −630 | 0.1012 | −632 |
| | 2.04 | In(III) | 0.143 | −598 | 0.1443 | −597 |
| | | Cd(II) | 0.190 | −638 | 0.1852 | −639 |
| | 4.09 | In(III) | 0.143 | −598 | 0.1498 | −594 |
| | | Cd(II) | 0.363 | −638 | 0.3722 | −637 |
| (B) Square wave polarography at the DME and square wave voltammetry at the HMDE | | | | | | |
| DME | 0.511 | In(III) | 0.550 | −592 | 0.5522 | −593 |
| | | Cd(II) | 0.165 | −636 | 0.1693 | −634 |
| | 2.04 | In(III) | 0.276 | −592 | 0.2804 | −595 |
| | | Cd(II) | 0.316 | −636 | 0.3082 | −637 |
| | 4.09 | In(III) | 0.276 | −592 | 0.2829 | −591 |
| | | Cd(II) | 0.621 | −636 | 0.6202 | −634 |
| HMDE | 0.511 | In(III) | 0.282 | −596 | 0.2723 | −596 |
| | | Cd(II) | 0.0886 | −640 | 0.09371 | −638 |
| | 2.04 | In(III) | 0.140 | −596 | 0.1397 | −596 |
| | | Cd(II) | 0.190 | −640 | 0.1825 | −641 |
| | 4.09 | In(III) | 0.140 | −596 | 0.1427 | −594 |
| | | Cd(II) | 0.361 | −640 | 0.3731 | −638 |

[a] Experimental parameters and definitions of terms are the same as in Table 5 except $c(Cd)/c(In)$: $1.78 \times 10^{-5}$ M/$3.48 \times 10^{-5}$ M = 0.511, $3.56 \times 10^{-5}$ M/$1.74 \times 10^{-5}$ M = 2.04, $7.12 \times 10^{-5}$ M/$1.74 \times 10^{-4}$ M = 4.09, $1.78 \times 10^{-4}$ M/$1.74 \times 10^{-5}$ M = 10.2.

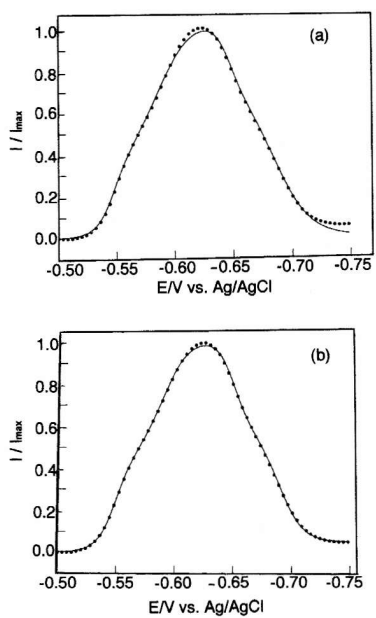

Fig. 6. Comparison of experimental (···) and calculated (solid line) differential pulse polarograms with $\Delta E = -100$ mV for reduction of a mixture of $1.74 \times 10^{-5}$ M In(III) and $3.56 \times 10^{-5}$ M Cd(II) in 1.0 M HCl. The reversible model was used assuming the absence of a baseline in (a) and with addition of a linear baseline in (b) for the curve fitting.

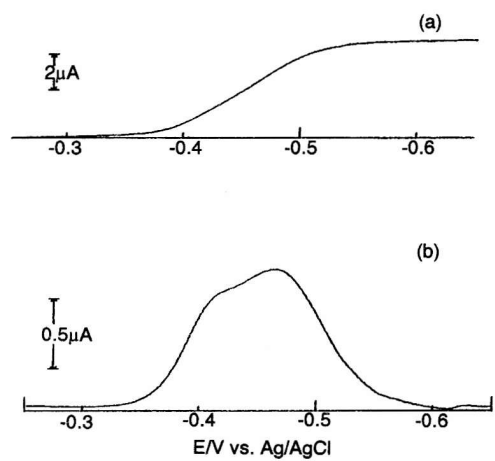

Fig. 7. Normal (a) and pseudo-derivative (b) pulse polarograms for the reduction of a mixture of $1.21 \times 10^{-4}$ M Pb(II) and $9.62 \times 10^{-4}$ M Tl(I) in 1.0 M ZnSO$_4$.

### Pb(II) + Tl(I) system

It is impossible to visually distinguish two responses from the normal pulse polarogram (Fig. 7a) for the reduction of the Pb(II) + Tl(I) system in the supporting electrolyte of 1.0 M ZnSO$_4$. However, two peaks can be detected in the pseudo-derivative normal pulse polarogram (Fig. 7b). Table 7 also shows that, as expected, the peak potentials for reduction of Pb(II) have different values in square-wave polarography, differential pulse polarography and pseudo-derivative normal pulse polarography. However, the calculated reversible half-wave potential of lead(II) at $-408$ mV in square wave polarography, $-409$ mV in differential pulse polarography and $-405$ mV in pseudo-derivative normal pulse polarography are almost technique independent as expected. An analogous situation applies for reduction of Tl(I).

Table 7
Resolution of overlapping processes for a mixture of $1.2 \times 10^{-4}$ M Pb(II) and $9.6 \times 10^{-4}$ M Tl(I) in 1 M ZnSO$_4$ using a range of voltammetric techniques [a]

| Technique | $\Delta E$ (mV) | Ion | Expected | | Calculated | |
|---|---|---|---|---|---|---|
| | | | $I^p$ ($\mu$A) | $E^p$ (mV) | $I^p$ ($\mu$A) | $E^p$ (mV) |
| PDNPP [a] | $-10$ | Pb(II) | 0.189 | $-400$ | 0.186 | $-400$ |
| | | Tl(I) | 0.404 | $-458$ | 0.409 | $-459$ |
| PDNPP [a] | $-50$ | Pb(II) | 0.742 | $-381$ | 0.689 | $-380$ |
| | | Tl(I) | 1.87 | $-437$ | 1.888 | $-438$ |
| PDNPP [a] | $-100$ | Pb(II) | 0.939 | $-356$ | 0.918 | $-360$ |
| | | Tl(I) | 3.01 | $-412$ | 3.077 | $-410$ |
| SWP | 5 | Pb(II) | 0.198 | $-408$ | 0.187 | $-408$ |
| | | Tl(I) | 0.429 | $-468$ | 0.440 | $-469$ |
| DPP [a] | $-10$ | Pb(II) | 0.184 | $-414$ | 0.192 | $-414$ |
| | | Tl(I) | 0.418 | $-474$ | 0.429 | $-474$ |

[a] Experimental parameters are: drop time, 1 s; pulse width, 0.06 s; current sampling time, 0.02 s; scan rate, 2 mV s$^{-1}$. Other experimental parameters and terms are defined in the text or in previous tables. PDNPP = pseudoderivative normal pulse polarography, SWP = square-wave polarography, DPP = differential pulse polarography.

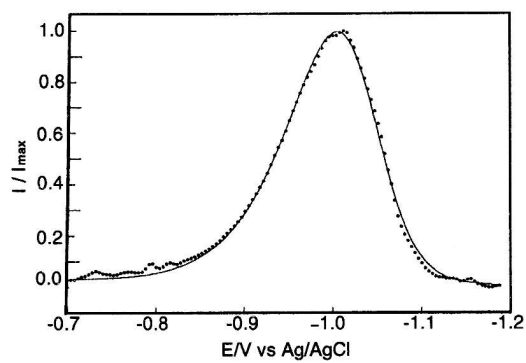

Fig. 8. Comparison of experimental (···) and calculated (solid line) pseudo-derivative normal pulse polarograms for reduction of $1.92 \times 10^{-4}$ M Cr(III) in 1.0 M NaCl.

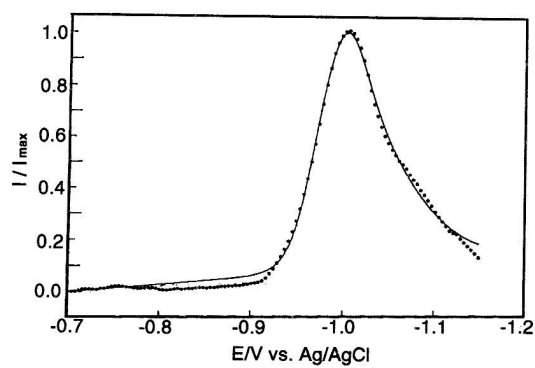

Fig. 9. Comparison of experimental (···) and calculated (solid line) pseudo-derivative normal pulse polarograms for reduction of $1.53 \times 10^{-4}$ M Zn(II) in 1.0 M NaCl.

### Cr(III) + Zn(II) system

In order to determine how well the irreversible model for Cr(III) and the quasireversible model for Zn(II) describe experimental curves, polarograms of the individual species were recorded, obtained under a range of conditions, and the fitting of the theoretical current–potential expression to the experimental curves was examined.

The fit of the irreversible model to the experimental pseudo-derivative normal pulse polarogram (Fig. 8) for the irreversible reduction of chromium ion in 1.0 M NaCl was excellent. The fitting of the quasireversible model to the experimentally obtained chromium(III) polarogram was equally good. However, the "goodness of fit" of the reversible model, successfully used above was much poorer than either the irreversible or quasireversible models perhaps because it has fewer adjustable parameters. The quasireversible model also may be used to describe the irreversible response because the irreversible reaction is a special case of the quasireversible reaction. However, the quasireversible peak model requires considerably longer computing time since it is more complicated and has more model parameters than the irreversible one, and therefore if the irreversible model is appropriate, it is still preferred.

Table 8
Resolution of overlapping processes for reduction of a mixture of $1.92 \times 10^{-4}$ M Cr(III) and $1.53 \times 10^{-4}$ M Zn(II) in 1 M NaCl with different pulse techniques and pulse amplitudes [a]

| Technique | $\Delta E$ (mV) | Ion | Expected | | Calculated | |
|---|---|---|---|---|---|---|
| | | | $I^p$ ($\mu$A) | $E^p$ (mV) | $I^p$ ($\mu$A) | $E^p$ (mV) |
| DPP | $-10$ | Cr(III) | 0.255 | $-966$ | 0.256 | $-962$ |
| | | Zn(II) | 0.406 | $-1020$ | 0.402 | $-1026$ |
| DPP | $-50$ | Cr(III) | 1.58 | $-954$ | 1.56 | $-949$ |
| | | Zn(II) | 1.67 | $-1004$ | 1.72 | $-1006$ |
| PDNPP | $-50$ | Cr(III) | 0.650 | $-1002$ | 0.675 | $-1000$ |
| | | Zn(II) | 0.395 | $-1020$ | 0.400 | $-1016$ |

[a] Other experimental parameters and terms are defined in the text or in previous tables.

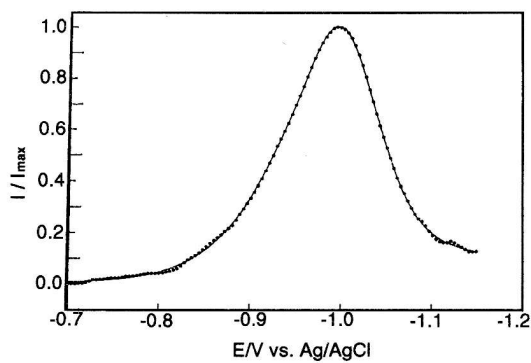

Fig. 10. Comparison of experimental (···) and calculated (solid line) pseudo-derivative normal pulse polarograms for a reduction of a mixture of $1.92 \times 10^{-4}$ M Cr(III) and $1.53 \times 10^{-4}$ M Zn(II) in 1.0 M NaCl.

The excellent fit of the quasireversible model to the experimentally obtained pseudo-derivative normal pulse polarograms for the reduction of zinc ion in 1.0 M NaCl confirms that this model also is suitable. The shoulder, observed at negative potentials in Fig. 9 has been observed in experimental differential pulse polarograms [34]. In the case of zinc, if the reversible or irreversible models are fitted, the relative standard deviation becomes much larger, consistent with the concept that the charge-transfer process is indeed quasireversible rather than reversible or irreversible. Fig. 10 demonstrates the overlapping response observed for a mixture of chromium and zinc. The digitally acquired experimental data and the corresponding model are in excellent agreement.

Data in Table 8 confirm that the resolution of the overlapping Cr(III) and Zn(II) peaks may be achieved over a wide range of pulse amplitudes and with either the pseudo-derivative normal pulse polarography or differential pulse polarography techniques. The relative standard deviations as the measurement of "goodness of fit" are very small in all cases.

## 6. Conclusions

The curve-fitting method described in this paper for resolving overlapped voltammetric processes achieves excellent resolution over very wide ranges of experimental factors (pulse amplitude, drop time, sweep rate, voltammetric technique, concentration ratio), and electrochemical parameters (electron numbers, peak shape, transfer coefficient, rate constant, peak separation). The method is robust, fast, and flexible, offering a number of advantages over other commonly used literature methods, in that it (a) is well-defined and statistically sound, (b) fully uses all the experimental data, and (c) has automatic initialization which removes subjective bias.